\documentclass[%
 aip,
 apl,
 sd,%
 amsmath,amssymb,
 reprint,%
]{revtex4-1}

\usepackage{siunitx}
\usepackage{graphicx}
\usepackage{dcolumn}
\usepackage{bm}

\begin{document}

\preprint{AIP/123-QED}

\title{Defect Line Terahertz Quantum Cascade Laser}

\author{A.~Klimont}

\author{R.~Degl'Innocenti}
 \email{rd448@cam.ac.uk}
 \affiliation{Cavendish Laboratory, University of Cambridge, J. J. Thomson Avenue, Cambridge CB3 0HE, United Kingdom}

\author{L.~Masini}
\affiliation{NEST, Istituto Nanoscienze-CNR and Scuola Normale Superiore, Piazza San Silvestro 12, 56127 Pisa, Italy}

\author{Y.~Wu}

\author{Y.~D.~Shah}

\author{Y.~Ren}
\altaffiliation[Current address: ]{Purple Mountain Observatory, Chinese Academy of Sciences, Nanjing 210008, China}

\author{D.~S.~Jessop}
\affiliation{Cavendish Laboratory, University of Cambridge, J. J. Thomson Avenue, Cambridge CB3 0HE, United Kingdom}

\author{A.~Tredicucci}
\affiliation{NEST, Istituto Nanoscienze-CNR and Scuola Normale Superiore, Piazza San Silvestro 12, 56127 Pisa, Italy}
\affiliation{Dipartimento di Fisica ``E. Fermi'', Universit\`a di Pisa, Largo Pontecorvo 3, 56127 Pisa, Italy}

\author{H.~E.~Beere}

\author{D.~A.~Ritchie}
\affiliation{Cavendish Laboratory, University of Cambridge, J. J. Thomson Avenue, Cambridge CB3 0HE, United Kingdom}

\date{\today}

\begin{abstract}
We present terahertz quantum cascade lasers operating at a defect mode of a photonic crystal bandgap. This class of devices exhibits single mode emission and low threshold current compared to standard metal-metal lasers. The mode selectivity is an intrinsic property of the chosen fabrication design. The lower lasing threshold effect, already reported in photonic crystal quantum cascade lasers, is further enhanced in the ultra-flat-dispersion defect line. The presented results pave the way for integrated circuitry operating in the terahertz regime and have important applications in the field of quantum cascade lasers, spectroscopy and microcavity lasers.
%
\end{abstract}

\maketitle

Quantum cascade lasers (QCLs) are well-established, compact semiconductor sources of coherent infrared and terahertz (THz) radiation. The fact that the vibrational and rotational resonances in many molecules lie in the THz region makes QCLs very attractive devices for applications in imaging\cite{Ren2015}, spectroscopy, and sensing\cite{Tonouchi2007}. These applications require a well defined single frequency mode of operation. However, typical THz QCLs exhibit a bandwidth of about \SI{200}{\giga\hertz}, which results in Fabry-P\'erot multi-mode lasing, as the ridge length is of the order of millimeters. Traditionally, single mode operation has been achieved by engineering frequency selective devices, such as distributed feedback resonators\cite{Mahler2004,Amanti2009}. This concept has been extended in two dimensions by implementing photonic crystals\cite{Colombelli2003a} (PhCs) for vertical emission, and for in-plane emitters by placing photonic crystals in front of laser facets\cite{Dunbar2005}. All these approaches increase the laser current threshold by introducing extra radiative channels. Keeping the threshold low and therefore achieving high wall-plug efficiency is crucial in limited electrical power environments, such as satellites, where THz QCLs serve as local oscillators in heterodyne receivers\cite{Hubers2005}.

Photonic crystals have proven to be a particularly interesting design concept for THz QCLs. Our approach is similar to what has been reported in Refs.~\onlinecite{Zhang2007, Benz2009, DeglInnocenti2016}. It is based on etching pillars in the active region and connecting them electrically with a metallic top layer. By using similar techniques, we demonstrate a series of defect line THz QCLs. The presented devices exhibit single frequency emission and low current threshold density $J_{th}$ compared to double metal (MM) QCLs of equivalent area which display multi-mode operation. Lasing occurs in the defects, and the frequency is fully tunable by modifying the size of the pillars. Our novel waveguide design can readily be extended to integrate QCLs in photonic circuits --- the defect line can be engineered to point in arbitrary directions or to include lossless bends. This is not possible in QCL ridges or omnidirectional PhCs. Finally, the defect line architecture is an excellent platform to study fundamental effects such as slow light or Purcell enhancement. To the best of our knowledge, this is the first demonstration of a defect line QCL.

\begin{figure}[b]
\includegraphics[scale=0.44]{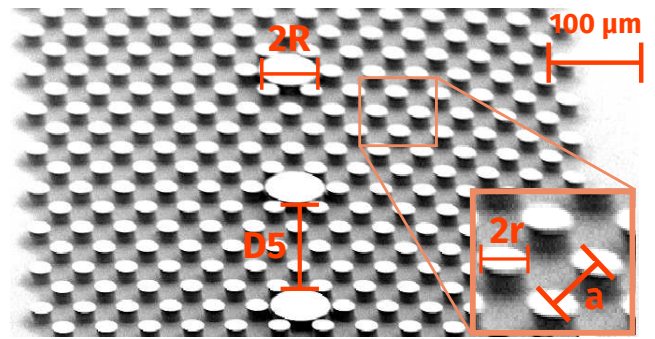}
\caption{\label{fig:sem} A SEM picture of the defect QCL after reactive ion etching and before BCB planarization. The pillars are $\sim$\SI{14}{\micro\meter} tall. The Au/Ni mask can be seen on top of the pillars.}
\end{figure}

In QCLs, intersubband selection rules define the outgoing light's polarization as TM. A lattice of high refractive index pillars in a low index medium gives rise to a TM photonic bandgap. We chose the triangular lattice, which results in the widest possible bandgap\cite{JoannopoulosBook}. The size of the gap depends on two factors: firstly, on the refractive index contrast between the pillars and the surrounding medium and secondly, on the ratio of pillar radius to the lattice constant ($r/a$, see Fig.~\ref{fig:sem}). The refractive index of the columns is defined by the active region material (GaAs/AlGaAs); we assumed the effective index $n_{\mathrm{AR}}=3.6$. We opted to planarize the structure with benzocyclobutene (BCB), which allows for a relatively simple fabrication while maintaining a significant contrast as $n_{\mathrm{BCB}}=1.55$. For these refractive indices, the maximum achievable relative band gap $\Delta \omega / \omega_0$ is $\sim0.3$ at $r/a = 0.25$, where $\omega_0$ is the frequency in the middle of the gap. For $\omega_0 = 2 \pi \cdot$\SI{2}{\tera\hertz}, this corresponds to the absolute gap width of \SI{0.6}{\tera\hertz}. Leaving air in between the pillars would offer a higher contrast and a bigger gap (\SI{1}{\tera \hertz}). This approach, although feasible\cite{Benz2009}, would potentially be less mechanically stable.

Bound-to-continuum QCL active region designs are known to have low thresholds and therefore were chosen for this work. A bound-to-continuum design with central frequency around \SI{2}{\tera\hertz}\cite{Worrall2006} was chosen because of continuous wave operation and low $J_{th} \simeq \SI{100}{\ampere \per \square \centi \meter}$. In order to place the middle of the gap around \SI{2}{\tera\hertz}, we set the lattice constant $a=42$--\SI{46}{\micro\meter}. The frequency range between \SI{2}{\tera\hertz} and \SI{3}{\tera\hertz} is particularly interesting for gas spectroscopy, as numerous molecules exhibit rotational and vibrational resonances in this frequency region. We decided to work on the lower frequency limit of this range due to less stringent fabrication accuracy requirements.

\begin{figure}[t!]
\includegraphics[scale=0.41]{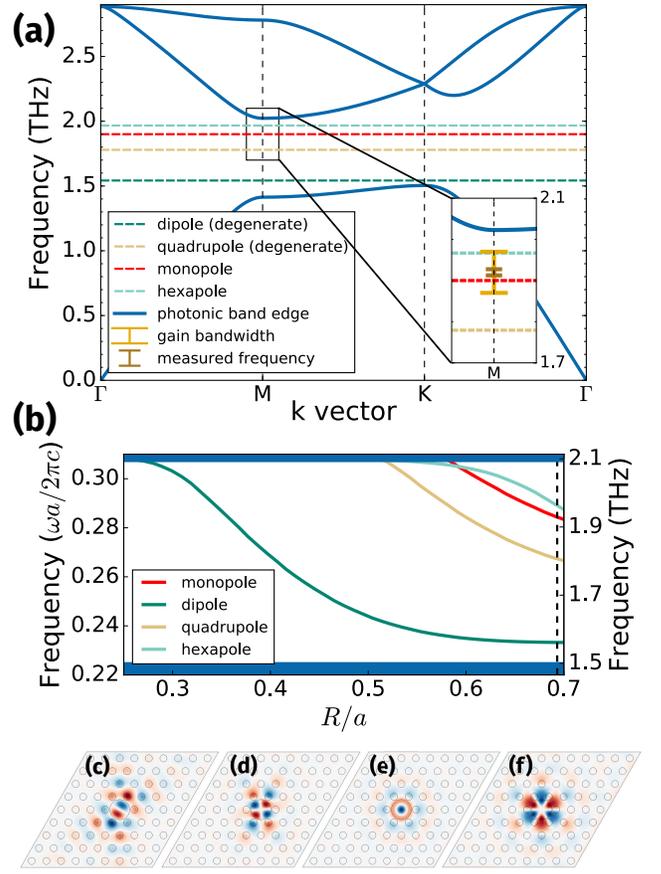}
\caption{\label{fig:bands} \textbf{a:} Photonic structure of the defect laser for the defect diameter $R=\SI{31}{\micro\meter}$, small pillar diameter $r=\SI{11}{\micro\meter}$, lattice constant $a=\SI{44}{\micro\meter}$ ($r/a=0.25$). Solid lines represent the air and dielectric band edges. The dashed lines between them show different modes allowed in the defect pillars. Inset: The yellow bracket spans the bandwidth of a reference MM device from the same active region. The brown bracket is shown at the measured emission frequency (taking into account FTIR resolution of \SI{0.25}{\per\centi\meter}), which is in good agreement with the calculation. The direction of the defect line corresponds to $\mathrm{\Gamma M}$ direction in reciprocal space. \textbf{b:} Frequency dependence of the defect modes on $R/a$. The horizontal bands on top and bottom represent the air and dielectric bands, respectively. The right axis shows absolute frequency for the chosen $a$. The dashed line marks the $R/a$ ratio of the simulation in the top figure. Electric field intensity are shown for \textbf{c:} dipole \textbf{d:} quadrupole \textbf{e:} monopole \textbf{f:} hexapole.}
\end{figure}

Introducing a series of defects in the photonic lattice allows for wave guiding, due to the conservation of crystal momentum\cite{JoannopoulosBook}. The defects pull down optical modes from the air band or push up modes from the dielectric band into the band gap, depending on whether their radius is larger or smaller compared to the surrounding pillars. A preferred monopole mode, whose electric field peaks in the defect, can be achieved by reducing the size of the pillars. This is unfortunate, as it reduces the amount of active gain material. However, for a large refractive index contrast and a certain range of $R/a$ ratio, larger defects can support higher order modes. For the chosen contrast (3.6:1.55) and defect radius (54--\SI{62}{\micro \meter}), four modes exist: dipole, quadrupole, monopole and hexapole (Fig.~\ref{fig:bands}c, d, e and f, respectively). The ``monopole'' mode is in fact a superposition of higher order modes. It has a strong electric field concentration in the defect, which makes it an ideal candidate for the lasing mode in a pillar-based defect line. The dipole and quadrupole are degenerate due to the symmetry of the crystal. The photonic band structure of the triangular lattice as well as the defect modes were calculated with MIT Photonic Bands software\cite{Johnson2001:mpb}. In order to simulate the defect line, we used a supercell (a defect pillar surrounded by four smaller pillars on each side) with periodic boundary conditions. Since the mode is confined vertically by two layers of metal, we simplified the problem to two dimensions, assuming infinite height of the pillars. The results of these simulations are shown in Fig.~\ref{fig:bands}.

In order to quantify the energy concentration of the modes in the active material, we define the overlap as

\begin{equation}
\Gamma = \frac{\int_{AR}\epsilon \Vert E \Vert^2}{\int \epsilon \Vert E \Vert^2}\,,
\end{equation}

where the integration in the numerator is over the pillars (active region) only, and the denominator is over the whole supercell, $\epsilon$ is the dielectric constant and $\Vert E \Vert$ is the norm of the electric field. $\Gamma$ for the monopole is approximately 0.89 and for the closest other mode --- hexapole --- it is only 0.7. Only these modes fall within the bandwidth of the active medium, therefore we assume that the monopole (Fig.~\ref{fig:bands}~c) is the emitted mode. It exists only in a short range of defect radii from $R=0.58a$ to $R=0.7a$ (bigger defects would touch the surrounding small pillars), which corresponds to \SI{25.5}{\micro\meter} and \SI{31}{\micro\meter}, respectively, for the chosen $a=\SI{44}{\micro\meter}$. The frequency of the monopole in this region ranges from \SI{2.08}{\tera\hertz} to \SI{1.94}{\tera\hertz}, respectively.

We fabricated the defect line QCLs from a bound-to-continuum active region emitting at \SI{2}{\tera\hertz}, with a bandwidth of $\sim$\SI{100}{\giga\hertz}\cite{Worrall2006}. It was wafer-wafer bonded to a GaAs substrate, which was subsequently polished and etched, as is done routinely for MM processing. A metal mask was defined by means of optical photolithography, and layers of Ti/Au/Ni (\SI{10}{\nano\meter}/\SI{500}{\nano\meter}/\SI{100}{\nano\meter}) were thermally evaporated. Nickel served as a sacrificial layer for the reactive ion etching (RIE) process. The $\sim$\SI{14}{\micro\meter}-tall pillars were etched in a JLS Designs RIE80 tool, using $\mathrm{SiCl_4:Ar}$ process gases in 6:10~sccm proportion. The etch rate was $\sim$\SI[per-mode=symbol]{100}{\nano\meter \per \minute}. Next, BCB was spun and cured, as described in Ref.~\onlinecite{DeglInnocenti2015}. In order to clean the top of the pillars from BCB and obtain a flat BCB surface around the pillars, the polymer was etched down chemically in $\mathrm{O_2:CHF_3}$ (30:20 sccm) atmosphere. Finally, the top Ti/Au (\SI{10/300}{\nano\meter}) contact was evaporated, the lasers were cleaved and mounted on copper blocks for efficient heat extraction. The reported devices were 0.5--\SI{0.7}{\milli \meter} long and $\sim$\SI{0.2}{\milli \meter} wide. The reference MM laser was cleaved into a ridge of a comparable area, $\SI{1}{\milli \meter} \times \SI{85}{\micro \meter}$.

\begin{figure}[t]
\includegraphics[scale=0.43]{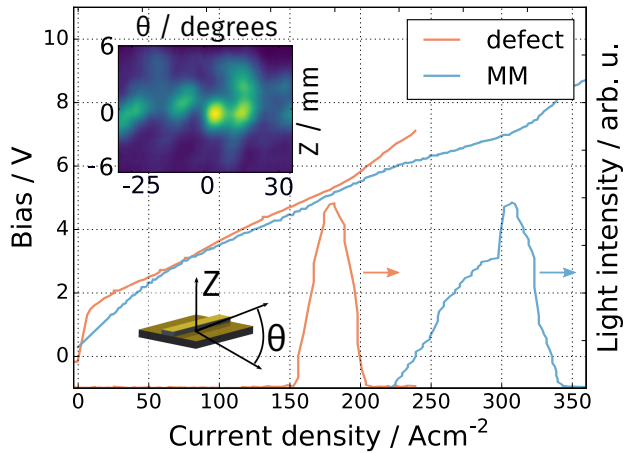}
\caption{\label{fig:iv_comparison} Light-current-voltage comparison between a defect and a MM device. Threshold current density estimated with this method is \SI{150}{\ampere \per \square \centi \meter}, which is 30\% lower than MM $J_{th}$. A much narrower dynamic range of the defect laser is an indication of single mode operation. The light intensity is not shown to scale, for clarity. \textbf{Inset:} Far-field emission pattern of the defect laser taken at \SI{18}{\milli\meter} from the facet. It is highly divergent due to interference and impedance mismatch at the facet, where active region pillars in a BCB matrix open into the free space.}
\end{figure}

\begin{figure}[t]
\includegraphics[scale=0.44]{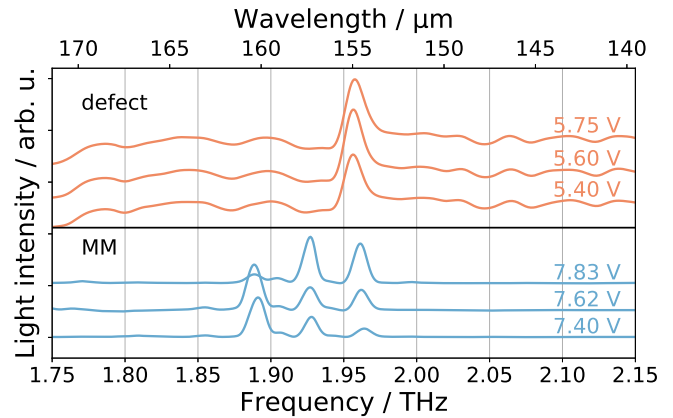}
\caption{\label{fig:spectra_comparison} Measured frequency spectra of a defect laser (top) and a reference MM device (bottom).}
\end{figure}

Here we report on four fully characterized devices with different defect radii (R) and distances between defects (D). Their performance is summarized in Table~\ref{tab:results}. Figures~\ref{fig:iv_comparison} and \ref{fig:spectra_comparison} show the results for one representative device (the top defect device in the table). We acquired the light-current-voltage (LIV) characteristics with a Tydex Golay cell and a lock-in amplifier gated at 9 Hz. The pulse generator that we used allowed us to test the laser with pulses from \SI{10}{\kilo\hertz} to \SI{150}{\kilo\hertz} and the duty cycle from 3\% to 70\% (pulse widths of \SI{200}{\nano \second} -- \SI{70}{\micro \second}) --- all devices lased throughout these ranges of parameters. The frequency spectra were acquired with a Bruker Fourier Transform Infrared spectrometer (FTIR), using a liquid helium cooled bolometer as the detector.

The light intensity curves already show signs of mode selectivity --- a narrower and more uniform dynamic range than in MM devices. This was confirmed by the FTIR spectra (Fig.~\ref{fig:spectra_comparison}). All of the fabricated defect QCLs were single mode, as expected from simulations. Moreover, by varying the size of the defect pillar, we were able to tune the emission frequency. The biggest defects gave rise to emission at \SI{1.89}{\tera\hertz}, whereas the smallest ones resulted in lasing at \SI{1.97}{\tera\hertz} (see Table~\ref{tab:results}). It should be noted that the monopole mode only exists within the photonic bandgap for a certain range of $R/a$ ratios, and the achieved emission frequencies are at the limits of this range. In order to select a mode outside of this range, the whole PhC structure (small pillars and defects) would have to be scaled. The only limitation for scaling is the resolution of photolithography. Covering the whole THz region (\SIrange{1}{10}{\tera\hertz}) only requires features with size of the order of micrometers, which do not pose technical challenges. We attribute the difference in bias of the compared lasers in Fig.~\ref{fig:spectra_comparison} to the composition of the top contact as well as the complexity of electronic transport in the defect laser.

\begin{table*}[t]
\begin{ruledtabular}
\begin{tabular}{lrrrrrrrr}
Type & R (\SI{}{\micro \meter}) & r (\SI{}{\micro \meter}) & f (\SI{}{\tera \hertz}) & $J_{th}$ (area) (\SI{}{\ampere \per \square \centi \meter}) & $J_{th}$ (IV) (\SI{}{\ampere \per \square \centi \meter}) & $J_{peak}$ (area) (\SI{}{\ampere \per \square \centi \meter}) & $T_{max}$ (K)\\
\hline
MM &  & &  1.88 -- 1.96 & 220 & & 310 & 75\\
D1 & $27 \pm 1$ & $10 \pm 1$ & 1.97 & 130 & 190 & 140 & 50\\
D1 & $31 \pm 1$ & $11 \pm 1$ & 1.89 & 230 & 205 & 270 & 60\\
D5 & $28.5 \pm 1$ & $10.5 \pm 1$ & 1.92 & 135 & 160 & 165 & 65\\
D5 & $28.5 \pm 1$ & $11 \pm 1$ & 1.96 & 100 & 150 & 190 & 50\\
\end{tabular}
\end{ruledtabular}
\caption{\label{tab:results} Measured properties of four different defect line lasers and a reference MM device. Higher emission frequencies are consistent with lower defect radii, which were measured with an SEM. We attribute the higher $J_{th}$ of the third device to the fact that the top contact did not completely cover the photonic crystal, giving rise to radiative loss from the top of the structure.}
\end{table*}

The calculation of $J_{th}$ in photonic crystal lasers is a non-trivial task\cite{DeglInnocenti2016}. Because BCB is an electrical insulator, the current only flows through the pillars. However, the non-uniform concentration of the optical mode in the defect affects electronic transport. The authors of Ref.~\onlinecite{Zhang2007} estimated $J_{th}$ by aligning current-voltage (IV) curves of a PhC device with a metal-metal (MM) reference laser. The active region that they used exhibited a pronounced feature in pre-threshold transport which served as the aligning point. The bound-to-continuum active region that we used did not show similar features. Instead, we aligned the slope of the defect and the reference MM IV curve pre-threshold. Using this method on a representative defect device, we calculated $J_{th}=\SI{160}{\ampere \per \square \centi \meter}$. This figure is 30\% lower than the reference MM device, where $J_{th}=\SI{220}{\ampere \per \square \centi \meter}$. This method of threshold estimation is illustrated in Fig.~\ref{fig:iv_comparison} --- the MM data are plotted in their raw form, but the defect results were artificially shifted in order to match the pre-threshold slope of the two IV curves.
The alternative and perhaps more intuitive way to calculate current density is to calculate the area of all pillars covered by the top Ti/Au contact and to divide the driving current (read from a current probe) by this value. The results are commensurate with the first method, and for the same representative device we obtained $J_{th}=\SI{130}{\ampere \per \square \centi \meter}$, corresponding to a 40\% reduction compared to the MM reference.

The maximum operating temperature was only slightly lower than that of the reference MM device ($T_{\mathrm{max (defect)}}=$\SIrange[range-phrase=--,range-units=single]{50}{65}{\kelvin} vs. $T_{\mathrm{max (MM)}}=\SI{75}{\kelvin}$). We attribute this effect to the BCB having a lower thermal conductivity than GaAs, which hampers heat extraction from the pillars.

Defect line QCLs have a much smaller effective lasing volume than ridge lasers --- defect pillars only take about 10\% of the device. Because of this, the peak output power from defect devices was an order of magnitude lower than that of MM lasers ($P_{\mathrm{defect}}=~$\SIrange[range-phrase=--,range-units=single]{0.45}{23}{\micro\watt} vs. $P_{\mathrm{MM}}=\SI{150}{\micro\watt}$). This could be addressed by increasing the density of defect pillars, i.e. reducing the distance between consecutive defects in line.

The theoretical framework to describe defect line resonators was laid out by Yariv and others\cite{Yariv1999}. The group velocity $v_g$ of light propagating in the big pillars is proportional to the coupling strength of adjacent defect cavities. Reducing $v_g$ increases the effective refractive index, which in turn leads to a higher electromagnetic field concentration and enhances the gain. $J_{th}$ should decrease proportionally to the group velocity. We propose this so--called slow light effect as the main mechanism responsible for $J_{th}$ reduction in our devices. In hope of quantifying this effect, we fabricated lasers with different distances between defect pillars. According to simulations, the dispersion of the defect modes ranges from ultra-flat for weakly coupled (D5, five rows of small pillars in between consecutive defects, as in Fig.~\ref{fig:sem}) to steep for highly coupled (D1, one row) defects. However, with the few devices that we characterized, we did not see a correlation between defect coupling and $J_{th}$. In order to better quantify this effect, more precise fabrication would be required, as small imperfections are detrimental to achieving ultra-flat dispersion.

Another effect which may play a role in $J_{th}$ reduction is Purcell enhancement, which modifies the local density of states of photons in a microcavity. This phenomenon has been proposed as a way to cut radiative emission lifetimes and therefore decrease $J_{th}$\cite{Walther2011}. The magnitude of the effect is proportional to $\frac{(\lambda/n)^3}{V}$, where $\lambda$ is the free space wavelength of emitted light, $n$ is the refractive index of the material, and $V$ is the volume of the laser cavity. For one defect pillar and the wavelength of \SI{150}{\micro\meter}, this ratio is around 2, which suggests a measurable reduction of lifetimes. The shortest device that we characterised comprised only three defects, but it is entirely possible to fabricate a single-defect microcavity. The exploration of the Purcell effect is beyond the scope of this article.

The demonstration of the defect line opens up a range of interesting possibilities in waveguide design. Because the small pillars around the defect form a lossless mirror, it would be possible to fabricate curved defect lines\cite{Yariv1999}. This would lift the long-standing constraint of QCLs emitting in-plane, namely rectangular ridges. Such curved devices could be more easily integrated in THz circuitry.

In summary, we presented a defect line THz QCL. Its advantages include single mode operation, low threshold current density, directionality and tunability. The results of our experiments are in excellent agreement with PhC theory. This waveguide architecture serves as an attractive platform to study the fundamental phenomena responsible for threshold reduction, namely the slow light and Purcell effects. Defect line QCLs are ideal candidates for THz applications where high wall-plug efficiency and mode selectivity is crucial.

The authors acknowledge the financial support by the Engineering and Physical Sciences Research Council, Grant EP/J017671/1 (CoTS) and by the European Research Council, Grant ERC-FP7-321122 (SouLMan).

\nocite{*}
\bibliography{defect1,defect2}

\end{document}